# Microresonator Frequency Comb Based High-Speed Transmission of Intensity Modulated Direct Detection Data

Peng Xing[1], George F. R. Chen[1], Hongwei Gao[1], Anuradha M. Agarwal[3,4], Lionel C. Kimerling[4,5], and Dawn T. H. Tan[1,2,*]

[1]Photonics Devices and System Group, Singapore University of Technology and Design, 8 Somapah Rd, Singapore 487372, Singapore

[2]Institute of Microelectronics, A*STAR, 2 Fusionopolis Way, #08-02, Innovis Tower, Singapore 138634, Singapore

[3]Microphotonics Center, Massachusetts Institute of Technology, 77 Massachusetts Avenue, Cambridge, Massachusetts 02139, USA

[4]Materials Research Laboratory, Massachusetts Institute of Technology, 77 Massachusetts Avenue, Cambridge, Massachusetts 02139, USA

[5]Department of Materials Science and Engineering, Massachusetts Institute of Technology, 77 Massachusetts Avenue, Cambridge, Massachusetts 02139, USA

*Corresponding author: dawn_tan@sutd.edu.sg

**Abstract**

Globally, the long-haul transmission of ultra-high bandwidth data is enabled through coherent communications. Driven by the rapid pace of growth in interconnectivity over the last decade, long-haul data transmission has reached capacities on the order of tens to hundreds of terabits per second, over fiber reaches which may span thousands of kilometers. Data center communications however operate in a different regime, featuring shorter reaches and characterized as being more cost and power sensitive. While integrated microresonator frequency combs are poised to revolutionize light sources used for high-speed data transmission over fiber, their use has been limited to coherent detection schemes. In this paper, we demonstrate the use of microresonator frequency combs pumped with a single laser for the transmission of high-speed data, importantly using direct detection schemes. We achieve 120 Gb s$^{-1}$ and 240 Gb s$^{-1}$ aggregate data transmission for 30 Gb s$^{-1}$ non-return-to-zero (NRZ) and 60 Gb s$^{-1}$ pulse modulation amplitude 4 (PAM4) modulation formats respectively over 2 km of optical fiber, exceeding the reach, single lane data rate, and aggregate data rates specified in Parallel Single Mode 4 (PSM4) and Course Wavelength Division Multiplex 4 (CWDM4) multi-source agreements. Remarkably, we achieve an extremely low power penalty of 0.1 dB compared to back-to-back characterization. The results firmly cement CMOS-compatible micro-resonator frequency combs based high-speed data transmission as a viable technology for the cost and power sensitive data center transceiver industry.

**Introduction**

Optical frequency combs have accelerated innovations in various fields [1-6] such as quantum and ultrafast optical information processing [7-10], microwave photonics [11, 12] and precision metrology [13-15]. In the commercially burgeoning field of silicon photonics-based data center communications, integrated optical frequency combs appear well poised to catalyze transformation. The promise of using a single laser to generate multiple wavelengths of light, each of which may serve as a vessel for data transmission is appealing both from a technological and cost standpoint [16-19]. Noteworthy applications of frequency combs for ultra-high density optical communications over fiber include tens of Tb s$^{-1}$ in Hydex microresonator combs recently reported by Moss et. al (44.1 Tb s$^{-1}$) [20], and silicon nitride microresonator combs (50 Tb s$^{-1}$) [21]. Non microresonator-based frequency combs generated through self-phase modulation in waveguides have enabled 661 Tb s$^{-1}$ of data transmission in multi-core fiber [22]. These impressive demonstrations of ultra-high-capacity data transmission powered by frequency combs point unequivocally to their developmental trajectory and potential to revolutionize telecommunications and data communications hardware architectures. In these aforementioned demonstrations of high-capacity data transmission powered by microresonator frequency combs, soliton states such as soliton crystals and dissipative Kerr solitons (DKS) were used to drive information transmission. Soliton crystal states are created with slow scans of the pump laser wavelength to transition from a chaotic state to the crystal state, both of which possess similar intracavity optical power [17, 28, 29]. In contrast, DKS states which are characterized by the soliton steps observed during laser scans perform optimally when the soliton state is achieved while the resonator is close to thermal equilibrium [18, 30 – 32].

Despite their promising performance demonstrated to date, integrated frequency combs for high bandwidth optical communications over fiber thus far have relied on coherent modulation formats and coherent detection [20 – 23]. The reliance of coherent communications on digital signal processing introduces a cost, power, complexity and latency overhead compared to intensity modulated direct detection (IMDD), with the complexity of digital signal processing adopted scaling with the fiber reaches [33]. Despite the promise of integrated microcombs for high-speed optical communications, the data center industry remains extremely cost and power sensitive; The commercial relevance of IMDD vs. coherent detection is further reflected in industry standards released by multi-source agreements (MSAs) including the Parallel Single Mode 4 (PSM4) [24], Course Wavelength Division Multiplexing 4 (CWDM4) [25], 100G Lambda [26] MSAs and the IEEE P802.3ba 100GBase standards [27], all of which utilize modulation schemes which work with direct detection. In many commercially deployed data center communication hardware, both within and outside of these MSAs, the reaches and specific applications served have a strong preference for IMDD from a cost perspective. Industry road maps describe the use of IMDD for shorter reaches, whereas coherent detection becomes viable only at much longer reaches and data rates on the order of hundreds of Gbaud s$^{-1}$ [33-35]. It remains to be seen, whether microresonator frequency combs can indeed supply the optical quality required for low bit error rates in IMDD-based high speed data transmission over fiber lengths deployed for data center communications, when the complex error correction functions associated with coherent detection are not available.

In this paper, we demonstrate the use of a silicon nitride microresonator for the transmission of 30 Gb s$^{-1}$ NRZ and 60 Gb s$^{-1}$ PAM4 IMDD data. We find that the primary comb state, also known as a Turing pattern which is easily pumped without requiring feedback or active stabilization, while also requiring a low pump power of 20 mW in our devices, were suitable for high-speed data transmission using direct detection, thus providing a promising, easily accessible, alternative means of high-speed data transmission other than soliton crystal or DKS states. In this reported regime, the comb state is sufficiently far from where the comb line output is susceptible to rapid fluctuations, such that any thermal fluctuations either from changes in intracavity power as the laser pump wavelength is tuned into the resonance or from ambient temperature will have negligible impact on the comb power stability. In several other comb states studied here, it was found that low bit error rates or open eye diagrams could not be obtained. These comb states corresponded to the regime in which the pump-resonance detuning was further decreased compared to the Turing state where the comb line amplitudes in contrast are defined by rapid fluctuations [30]. Importantly, we demonstrate the use of high-speed data transmission using an integrated frequency comb, using industrially pertinent intensity modulated direct detection, as opposed to coherent detection. We obtain open eye diagrams for the transmission of data over 2 km of optical fiber, which meets or exceeds the reach specified in both PSM4 and CWDM4 MSAs. Bit error rates smaller than the forward error correction (FEC) limit for NRZ data are obtained for the various characterized comb lines. Using a microresonator pumped with a single laser, we achieve an excess of 120 Gb s$^{-1}$ and 240 Gb s$^{-1}$ aggregate data transmission form NRZ and PAM4 modulation formats, exceeding both the single lane and aggregate data rates specified in the PSM4 and CWDM4 MSAs. An extremely low power penalty of only 0.1 dB compared to back-to-back characterization is achieved. The results represent high-speed transmission of IMDD data at baud rates exceeding that used in both PSM4 and CWDM4 transceivers.

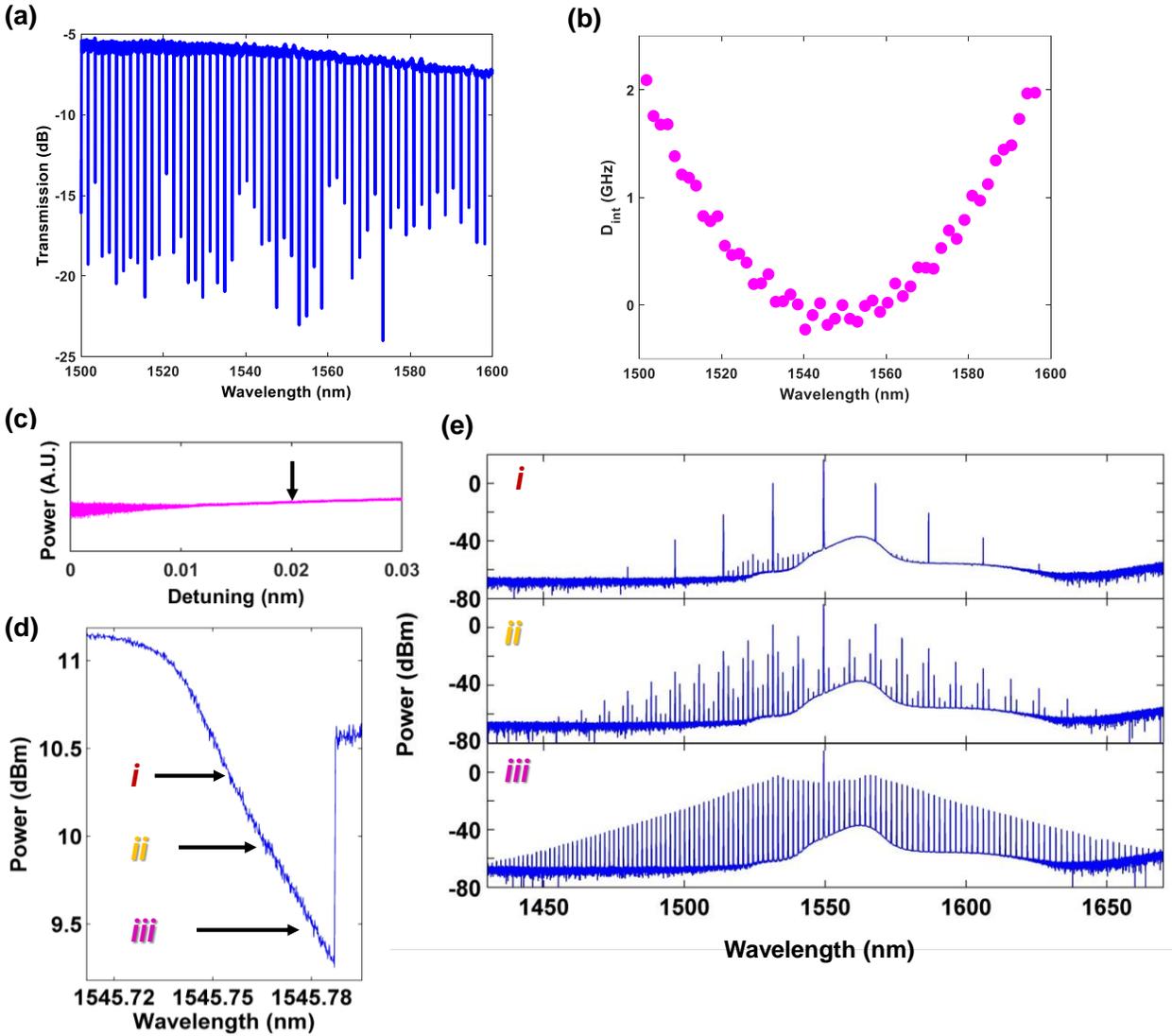

Figure 1. Silicon nitride microresonator and frequency comb properties. The different pump regimes and frequency comb output are described. (a) Experimentally measured transmission spectrum and (b) dispersion of the microresonator. (c) Measured output intensity of the frequency comb as a function of pump-resonator detuning. The arrow indicates the pump regime used in the experiments. (d) Measured transmission of the resonance as the laser wavelength is swept. The locations of the pump wavelength used to excite comb states *i*, *ii* and *iii* are denoted. (e) Experimentally generated frequency comb states corresponding to the pump wavelengths shown in Fig. 1 (d).

**Results**

We utilize a silicon nitride microresonator with a loaded and intrinsic quality factor of $1.2 \times 10^6$ and $1.9 \times 10^6$ respectively. Figures 1 (a) and (b) show the measured transmission spectrum and dispersion profile of the resonator. The wavelength of the pump resulted in several different comb states, as shown in Fig. 1 (d) and (e). The frequency comb states shown were generated by sweeping a continuous-wave laser from the blue side into a preset wavelength within the resonance. These different present wavelengths give rise to the different comb states as shown in Fig. 1 (e). Figure 1 (c) shows the experimentally measured comb output as a function of the pump-resonance wavelength detuning. It is observed that at the detuning used in our high-speed experiments (indicated by a black arrow), the transmitted power has minimal fluctuations, and does not experience abrupt changes in amplitude when the detuning value is either increased or decreased. This is of profound significance, as it implies that the amount of intracavity power circulating in the microresonator is not subject to large fluctuations, rendering the comb state less susceptible to destabilization. Conversely, Fig. 1 (c) shows that when the pump-resonance detuning is decreased beyond 0.01 nm, the transmitted power fluctuates considerably with marginal wavelength changes, and is likewise not ideal for a stable output amplitude that is intrinsically required for low IMDD bit error rates. These prior observations on unstable, fluctuating comb line amplitude were corroborated in our experiments through poor eye diagrams obtained in comb states *ii* and *iii*.

Comb state *i* shown in Fig. 1 (d) represents the pump regime for the primary comb state which we adopt for our high-speed data experiments. In modulation schemes compatible with IMDD such as NRZ and PAM4, the optical carrier undergoes intensity modulation, with NRZ and PAM4 possessing two and four different amplitude levels respectively. It is thus imperative that the output from the comb lines need to have a stable amplitude as a function of time.

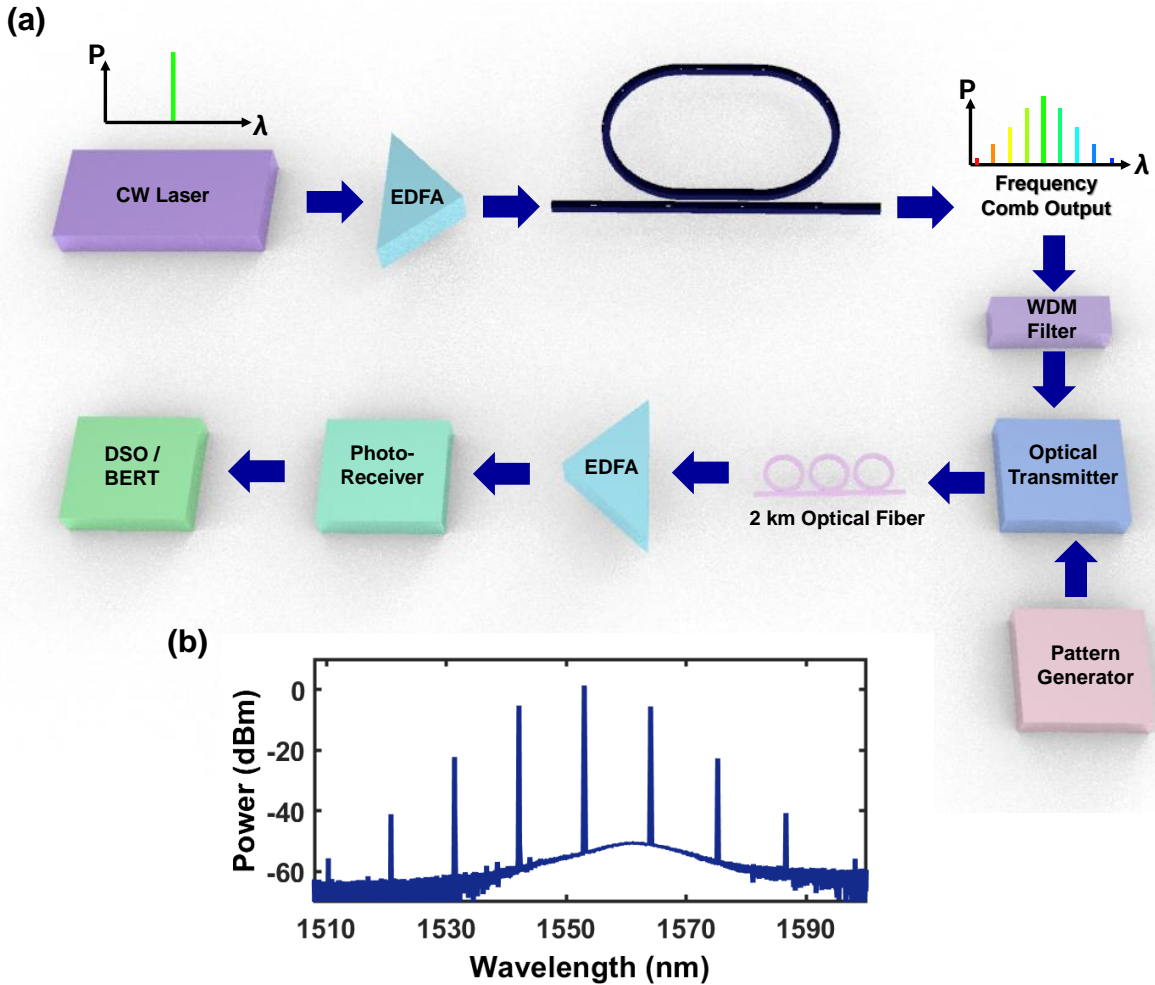

Figure 2. Experimental setup and frequency comb state used for the high-speed data experiments. (a) Schematic of the IMDD high-speed data (30 Gb s$^{-1}$ NRZ and 60 Gb s$^{-1}$ PAM4) experiments using the microresonator frequency comb. (b) Spectrum of the comb state used for the high-speed data experiments. The pump wavelength and power are 1553 nm and 20 mW respectively. DSO – Digital Sampling Oscilloscope, BERT – Bit Error Rate Tester

Figure 2 (a) shows the schematic of the experimental setup used. We characterize the light at the pump wavelength, and three next nearest neighbor comb lines. These span from a wavelength of 1531 nm to 1565 nm. We note that the second nearest comb line on the red side of the pump at 1576 nm could not be characterized as it was outside the range of the EDFA preceding the photoreceiver. However, it has the for 30 Gb s$^{-1}$ NRZ data and 60 Gb s$^{-1}$ PAM4 data. The characterization is performed for the pump wavelength (1553 nm) and comb lines at 1531 nm, 1542 nm and 1564 nm.

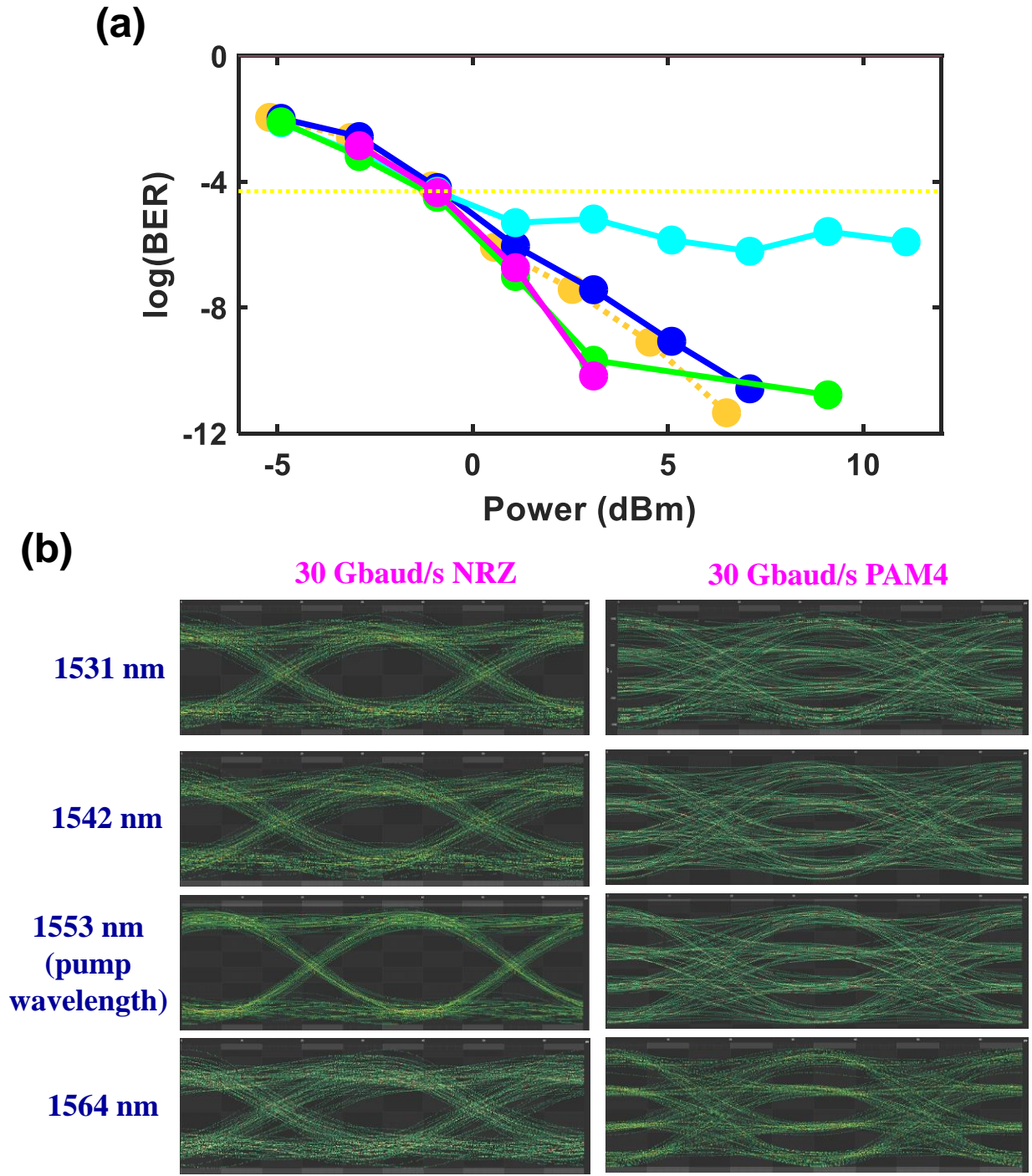

Figure 3. High-speed data experimental characterization showing BER less than the FEC threshold and open eye diagrams for all four comb lines. (a) Measured bit error rate for 30 Gb s$^{-1}$ NRZ data for the pump wavelength of 1553 nm (blue) and comb lines at 1531 nm (cyan), 1542 nm (fuchsia), 1564 nm (green) and B2B (orange dashed line). The FEC limit of $5 \times 10^{-5}$ is shown as the yellow dashed line. (b) eye diagrams

same power as the comb line at 1542 nm and is likely to have similar bit error rate (BER) and eye diagram properties. First, the frequency comb is generated by pumping the resonance at 1553 nm using a pump power of 20 mW. The generated comb lines are then encoded with 30 Gb s$^{-1}$ NRZ and 60 Gb s$^{-1}$ PAM4 data. The high-speed data is then transmitted through a 2 km long optical. This fiber length is especially industry relevant, given that PSM4 specifications point to a maximum reach of 500 m and CWDM4 specifications point to a 2 km maximum reach [24, 25]. Details of the high-speed characterization are provided in Materials and Methods. Figure 3 shows the BER for the pump wavelength (1553 nm), and comb lines located at 1531 nm, 1542 nm and 1565 nm. For all characterized lines, we obtain a BER that is lower than the FEC limit (BER = $5 \times 10^{-5}$). The eye diagrams measured using the digital sampling oscilloscope further shows that the eye diagrams for both NRZ and PAM4 data are open, with the 2-level and 4-level amplitudes distinct. These results showcase the frequency comb output's suitability for serving as a vessel for IMDD high-speed data, importantly not requiring the use of expensive and complex coherent receivers. We analyze the power penalty of the frequency comb lines compared to back-to-back characterization at the FEC limit. The maximum power penalty of only 0.1 dB occurs for the comb line located at 1531 nm, which is located two lines away from the pump, illustrating the excellent optical quality and stability of the generated comb lines.

**Discussion**

We distinguish our work from prior work in high-speed data transmission over fiber using microresonator frequency combs, by demonstrating direct detect, intensity modulated data at data rates and reaches exceeding those in commercially deployed data center transceiver products, as opposed to coherent modulation formats and coherent detection, thus firmly cementing the applicability of frequency comb-based light sources as a promising alternative to the status quo which utilizes one laser per wavelength channel of data. We note that the comb state used for the high-speed data transmission did not require any active stabilization, and the comb output was stable because the pump-resonator detuning used, (i) resided in the regime where intensity fluctuations were absent in the comb lines and (ii) was in a regime of good thermal (wavelength) equilibrium. The output amplitude profile corresponding to the lines of specific comb states is another important determinant as to whether IMDD data may successfully use these generated wavelengths as a transmission vessel. DKS combs have been documented to temporally output different numbers of solitons, with the repetition rate of the solitons being related to the free-spectral range of the microresonator. Similar mechanisms also exist in soliton crystal states, with a taxonomy of soliton crystal states describing the frequency response of the generated comb and the corresponding number of solitons having previously been documented [28]. Considering the reliance of IMDD data on a constant, non-fluctuating output amplitude, soliton-states in high-speed data transmission are best used when the CW components in each comb line are filtered for use as individual data carriers, as has been done remarkably in Refs 20 and 21. In the context of wavelength multiplexed IMDD-based transceivers in the data center industry, wavelength spacing however needs to be much wider (on the order of 3 THz for CWDM) than the typical line spacing on the order of tens to hundreds of GHz in state-of-the-art soliton microresonator frequency combs. In our work, the comb spacing achieved was 1.4 THz, which is very close to the requisite wavelength spacing as set out by ITU CWDM requirements adopted by CWDM4-based transceiver products.

|  | **PSM4 SPECIFICATIONS** | **CWDM4 SPECIFICATIONS** | **OUR FREQUENCY COMB SOURCE** |
| --- | --- | --- | --- |
| **REACH** | Up to 500 m | 2 m – 2 km | 2 km demonstrated |
| **SINGLE LANE BAUD RATE** | 25 Gbaud s$^{-1}$ | 25 Gbaud s$^{-1}$ | 30 Gbaud s$^{-1}$ |
| **BIT ERROR RATE REQUIREMENT** | 5 x 10$^{-5}$ | 5 x 10$^{-5}$ | < 10$^{-5}$ |
| **SIDE MODE SUPPRESSION RATIO (MIN)** | 30 dB | 30 dB | > 35 dB |

Table 1: Comparison of the frequency comb source for IMDD high-speed data, with transmitter specifications for the PSM4 and CWDM4 MSAs. The metrics achieved in our frequency comb source meet or exceed those stipulated by the MSAs.

Table 1 shows the key transmitter specifications for products under the CWDM4 and PSM4 MSAs. In our work, we successfully demonstrate a microresonator-based frequency comb for the transmission of high-speed NRZ and PAM4 data at 30 Gbaud s$^{-1}$. This exceeds the baud rate used by companies delivering silicon photonics-based transceiver products: CWDM4 and PSM4 MSAs for example specify a baud rate (per lane) of 25 Gbaud s$^{-1}$ [24, 25, 35, 37]. In addition, the side mode suppression ratio which measures the difference in amplitude between the main laser mode and any side modes is specified as 30 dB for both PSM4 and CWDM4. In our frequency comb source, the side mode suppression ratio for each of the characterized frequency comb lines exceeds 35 dB. We note that IMDD based transceiver products typically adopt the O-band. A key reason being that chromatic dispersion in single-mode fiber is lowest at 1310 nm, and the 4 CWDM channels as specified by the international telecommunications union (1271 nm, 1291 nm, 1311 nm and 1331 nm) have substantially smaller dispersion than at the C- and L-bands [38]. The limitations imposed by dispersion impairments have been highlighted by various companies [39–42]. While coherent detectors can ameliorate impairments from dispersion, transceivers which adopt coherent detection technologies are typically only deployed at long reaches and data rates on the order of hundreds of Gb s$^{-1}$, primarily because the associated cost, complexity and latency overheads are prohibitive for short to mid-haul transmission reaches [33-35]. While our frequency comb technology has been demonstrated at the C- and L-bands, it may easily be translated to the O-band by tailoring the pump wavelength and/or dispersion profile of the resonator for that wavelength. Despite the higher optical fiber dispersion at the wavelengths used in this work (16 ps nm$^{-1}$ at 1550 nm vs. close to 0 ps nm$^{-1}$ at 1310 nm), we note that the quality of the received IMDD data showed low bit error rates (below the FEC limit of $5 \times 10^{-5}$), a marginal 0.1 dB power penalty compared to B2B characterization, and open eye diagrams. This further illustrates that within the 2 km fiber length used in this work, which is the maximum reach specified by the CWDM4 MSA (2 m – 2 km) and exceeds that specified by the PSM4 MSA (2 m – 500 m), the microresonator frequency comb demonstrated allows direct detection high-speed data to be achieved at baud rates which exceed current industry standards. We envision that in the future, the use of CMOS-compatible, homogeneously integrated on-chip dispersion compensation devices similar to that reported in Ref. 43 could avail far longer reaches without dispersion impairments, using the same comb state while enabling a monotonic left-ward shift in the bit error rate plot shown in Fig. 3 (a). This work demonstrates the strong potential for microresonator frequency combs to replace IMDD transceiver architectures in the data center industry, which today use one laser per channel (off-chip or heterogeneously integrated), importantly at commercially relevant reaches and data rates.

**Acknowledgments**

Funding from the National Research Foundation Competitive Research Grant (NRF-CRP18-2017-03), Ministry of Education ACRF Tier 2 Grant and A*STAR grant is gratefully acknowledged.

**Conflict of Interest Statement**

The authors declare no conflict of interest.

**Materials and Methods**

**Microresonator Design**

The frequency comb was generated using a silicon nitride ring resonator with a radius of 100 μm and cross section of 800 nm by 1.5 μm with $SiO_2$ over- and under-cladding. The ring's free-spectral-range, loaded and intrinsic quality factors were of 230 GHz, 1.2 million and 1.6 million respectively. Inverse tapers were to facilitate fiber-waveguide coupling with a coupling loss of 3 dB per facet. $D_{int}$ was calculated using the resonance locations of the measured microresonator transmission spectrum using the equation, $D_{int} = \frac{D_2}{2!}\mu^2 + \frac{D_3}{3!}\mu^3 + \cdots$ where $\mu$ is the relative mode number ($\mu = 0$ at the pump wavelength), $D_2$ and $D_3$ are the 2$^{nd}$ and 3$^{rd}$ order dispersion parameter.

**Details of frequency comb generation**

The transmission spectrum of the microresonators was characterized using a wavelength swept, tunable continuous-wave laser and a synchronized photodetector. The light was first adjusted for transverse-electric polarization prior to coupling into the microresonator. The output intensity of the generated frequency comb as a function of pump-resonance detuning was measured using a fiber-coupled ultrafast photodetector. To generate the microresonator frequency comb state, the tunable laser was swept from the blue side of the resonance to the specified laser-resonance detuning wavelength, at a scan rate of 10 nm s$^{-1}$. An erbium doped fiber amplifier was used to amplify the tunable laser output. The light was adjusted for transverse-electric polarization before being coupled into the microresonator. No active thermal stabilization or temperature control was required during the frequency comb generation and high-speed experiments.

**High-speed data experiments**

High-speed data was encoded onto the individual comb lines generated in the microresonator. A Pseudo Random Binary Sequence of $2^{31}-1$ (PRBS31) was used in the generation of bit patterns. The modulation formats used were 30 Gb s$^{-1}$ Non-Return-Zero-On-Off-Keying (NRZ-OOK) and a 60 Gb s$^{-1}$ Pulse Amplitude Modulation 4 Levels (PAM4). A Mach Zehnder Optical Transmitter modulates a Continuous Wave (CW) Laser centered at the pump wavelength, using PRBS31 patterns generated by the Bit Error Rate Tester's (BERT) pattern generator. The high-speed data encoded on the frequency comb lines was transmitted through a 2 km long single mode fiber. At the end of the fiber, the output light is coupled into a photoreceiver for Optical-To-Electrical conversion. The converted electrical data signal was analyzed using a digital sampling oscilloscope (DSO) to extract its eye diagram. Bit error rate measurements were performed using the BERT Receiver. The high-speed characterization is performed against a Back-To-Back (B2B) setup whereby the microresonator is replaced by an optical attenuator with an equivalent insertion loss as the microresonator.